\documentstyle[preprint,aps]{revtex}
\begin{document}
\preprint{\today}
\title{\bf Dramatic Switching of Magnetic Exchange in a 
Classic Transition Metal Oxide: Evidence for Orbital Ordering}
\author{Wei Bao,$^{1,2}$\cite{BNL} C. Broholm,$^{1,3}$ G. Aeppli,$^{4}$
P. Dai,$^{5}$ J. M. Honig,$^{6}$ and P. Metcalf$^{6}$}
\address{$^{1}$Dept. of Physics and Astronomy, The Johns Hopkins University, Baltimore, MD 21218\\
$^{2}$Physics Dept., Brookhaven National Laboratory, Upton, NY 11973\\
$^{3}$National Institute of Standards and Technology, Gaithersburg, MD 20899\\
$^{4}$NEC, 4 Independence Way, Princeton, NJ 08540 \\
$^{5}$Solid State Division, Oak Ridge National Laboratory, Oak Ridge, TN 37831\\
$^{6}$Purdue University, West Lafayette, IN 47907}
\maketitle
\begin{abstract}
Spin correlations in metallic and insulating phases of $\rm V_2O_3$ and its 
derivatives are investigated using magnetic neutron scattering. Metallic 
samples have incommensurate spin correlations varying little
with hole doping. Paramagnetic insulating samples have
spin correlations only among near neighbors.
The transition from either of these
phases into the low temperature insulating antiferromagnetic phase
is accompanied by an abrupt change of dynamic magnetic short range order.
Our results support the idea that the transition into the antiferromagnetic
insulator is also an orbital ordering transition.
\end{abstract}

\narrowtext

Orbital order and disorder have been popular ingredients for explanations 
of many curious phenomena in solids, most notably the transition metal 
oxides\cite{tmo,castl}. 
In spite of large interest among condensed matter physicists, the evidence
for non-trivial, in the sense of not simply being due to conventional 
spin-orbit coupling, orbital fluctuations and order is generally quite indirect.
We have been able to make progress by taking advantage of a very 
sensitive measure of electronic orbitals, namely the exchange interactions
linking the spins on neighboring atoms, which are
determined by integrals over outer electron wave functions.
In particular, we report the discovery 
of a dramatic switching of magnetic exchange interactions on moving 
between the high and low temperature insulating phases of the classic 
transition metal oxide V$_2$O$_3$. The switching is most naturally 
explained in terms of an orbital ordering transition. In addition, 
we find that the high temperature spin fluctuations in metallic
V$_2$O$_3$ are insensitive to hole doping, showing little evolution on going
from a stoichiometric sample with an insulating antiferromagnetic 
ground state to a sample which remains metallic down to the lowest 
temperatures and displays a low moment spin density wave (SDW). Furthermore,
the magnetic fluctuations in the paramagnetic metal are similar to
those in the paramagnetic insulator,
which suggest
that orbital fluctuations play an important role in the metal as well as 
in the insulator.

Fig.~1 shows the phase diagram of $\rm V_2O_3$ as a function of 
temperature and two kinds of doping. Vanadium deficiency drives the
material metallic by populating the 3d band with holes, while Cr$^{3+}$
(3d$^3$) substitution for V$^{3+}$ (3d$^2$) stabilizes an insulating state\cite{phase_d}.
Three single crystal samples studied in this work, and indicated by
color bars on the figure, cover all four phases of V$_2$O$_3$.
The samples were grown using a skull
melter and in the stoichiometric and vanadium deficient samples, oxygen
content was controlled to within $\delta y$=0.003 by annealing sliced
crystals for two weeks at 1400$^o$C in a suitably chosen CO-CO$_2$
atmosphere\cite{growth}. The sensitivity of our experiment was
increased by mutually aligning several crystals so that the total mass
reached $\sim$15g for $\rm V_{1.973}O_3$, 8g for $\rm V_2O_3$ 
(N\'{e}el temperature T$_N$=170 K) and 6.4g in the case of $\rm
(V_{0.97}Cr_{0.03})_2O_3$ (T$_N$=180 K).

The inelastic neutron scattering experiments were carried out on
thermal neutron triple axis spectrometers at the NBSR and HFIR
of NIST and ORNL respectively.
A pyrolytic graphite filter was used to remove high-order neutrons.
After correcting for the $\hbar\omega$-dependent efficiency of the
spectrometer, the
magnetic neutron scattering intensity was normalized to inelastic
scattering from transverse
acoustic phonons to yield absolute measurements of the
dynamic spin correlation function\cite{Lovesey},
\begin{eqnarray}
{\cal S}({\bf Q},\omega )& =& \frac{1}{2}
\sum_{\alpha\beta}(\delta_{\alpha\beta}-\hat{ Q}_\alpha\hat{ Q}_\beta )
|F(Q)|^2 \frac{(g\mu_B )^2}{2\pi N\hbar } \nonumber \\
& & \int dt\ e^{i\omega t }\sum_{\bf RR^\prime}
 e^{-i{\bf Q}\cdot ({\bf R}-{\bf R}^\prime)}
\langle S_{\bf R }^\alpha (t) S_{\bf R ^\prime}^\beta (0) \rangle.
\end{eqnarray}
Wave-vector,
${\bf Q}$, will be indexed in the hexagonal reciprocal lattice with
$a^* = 4\pi/\sqrt{3}a = 1.47(1)\AA^{-1}$ and $c^* = 2\pi/c =
0.448(1)\AA^{-1}$ in the metallic phases, $a^*= 1.46(1)\AA^{-1}$ and
$c^*=0.449(2)\AA^{-1}$ in the AFI phase, and $a^*=
1.45(1)\AA^{-1}$ and $c^*=0.451(2)\AA^{-1}$ in the PI phase.

 We begin by surveying the spin fluctuations in three
of the phases of V$_2$O$_3$: the metallic antiferromagnet (SDW), 
the metallic paramagnet (PM), and the insulating paramagnet (PI) (Fig.~1). 
Other workers have successfully surveyed spin waves in the fourth phase, 
the insulating antiferromagnet (AFI)\cite{wordr}.
Fig.~2(a) and (b) show contour maps of the dynamic 
spin correlation function ${\cal S}({\bf Q},\omega)$ in
metallic antiferromagnetic V$_{1.973}$O$_3$ at 1.4K and in paramagnetic, 
but still metallic V$_2$O$_3$ at 200K. In both cases, we observe ridges,
with bandwidths exceeding 18meV and
centered near {\bf Q}=(1,0,$\overline{0.3}$) and (1,0,2.3), wave vectors 
which characterize the magnetic order in the hole-doped material (e.g.,
V$_{1.973}$O$_3$)\cite{our}. 
As is appropriate, given the higher temperature for frame
(b) than frame (a), the ridges are sharper for frame (a).
On moving to the PI phase (Fig.~2(c)) at nearly the
same temperature, however,
there is further broadening. Indeed, the data are now best described as 
a single broad ridge along the $\hbar\omega$-axis  centered at 
{\bf Q}=(1,0,0.8). From the 
half-width-at-half-maximum of constant energy cuts through this ridge
(see Fig.~3(d) and ref.~\cite{longp}),
we estimate spin correlation lengths $\xi_c \approx 1.5 \AA$ and $\xi_a 
\approx 2.0\AA$ along the c-axis and in the basal plane respectively.

One of the most remarkable features of the metal-insulator transition
in $\rm V_2O_3$ is that the antiferromagnetic order (see inset of Fig.~1)
 which develops in
the insulator is different from the SDW which occurs in vanadium
deficient samples\cite{our,newb}. Specifically, as may also be seen in the inset
 of Fig.~1,
vanadium atoms in $\rm V_2O_3$ have three nearest neighbors within a
puckered honeycomb plane. In the SDW phase each spin is approximately
antiparallel to all of its three in-plane neighbors, whereas in the AFI
the three-fold symmetry is broken with one nearest neighbor parallel,
the two others antiparallel. The two types of local spin arrangements
are conveniently labeled $(10\ell )$-type and $(0.5,0.5,\ell )$-type
respectively, according to which line in reciprocal space contains
the magnetic Bragg peaks of the corresponding long range ordered 
structure. In the following we show that irrespective of whether we
consider static or dynamic properties, $(0.5,0.5,\ell )$-type correlations
exist only in the AFI phase while $(1,0,\ell )$-type correlations
exist only outside this phase.

To probe dynamic correlations corresponding to the two
structures, we performed constant-energy scans
along each of these two directions in 
reciprocal space. 
The results are shown in Fig.~3 where the right and left 
columns probe $(10\ell )$-type and $(0.5,0.5,\ell )$-type 
correlations respectively.
Filled symbols correspond to high temperature and
open symbols to low temperature phases. For both $\rm V_{2-y}O_3$ 
(top frames) and $\rm V_{1.94}Cr_{0.06}O_3$ (bottom frames), 
$(10\ell )$-type correlations are visible only outside the AFI phase whereas 
$(0.5,0.5,\ell )$-type correlations can be seen 
only in the AFI phase. Entry to the AFI phase not only changes
the near neighbor correlations, it also brings about coherence
in the magnetic excitations as evidenced by the resolution-limited double peaks
in Fig.~3(a) and (c). These correspond to the excitation of counter-propagating spin waves in the long range ordered 
antiferromagnet\cite{wordr}. 
Such dramatic modifications of spin dynamics 
indicate that exchange interactions undergo profound changes
at the transition to the AFI. At the same time, doping to produce either
an insulating phase by Cr substitution or a more metallic sample by 
decreasing the V to O ratio has a much smaller effect on the spin dynamics
at fixed temperature within the paramagnetic phase. Indeed, Fig.~3(b)
shows that at 200K, the magnetic fluctuations in V$_2$O$_3$ and V$_{1.97}$O$_3$,
which have AFI and metallic SDW ground states respectively, are identical.

Fig.~4 gives the detailed temperature dependence of dynamic
spin correlations for $\rm V_2O_3$ and $\rm
V_{1.94}Cr_{0.06}O_3$. Coincident with the
transition to the AFI phase (vertical dashed lines) is an abrupt switch 
between
the two types of dynamic spin correlations. A
remarkable similarity exists between the transitions to the
AFI phase from the paramagnetic metal ($\rm V_2O_3$, left column)
and from the paramagnetic insulator ($\rm V_{1.94}Cr_{0.06}O_3$, right
column), which suggests a common mechanism
which is independent of whether or not a Fermi surface, associated with 
metallic behavior, exists in the high temperature phase.

Our discoveries find no comprehensive explanation within a one-band 
Hubbard model\cite{morii,rozb}. The most serious difficulty is
the abrupt switch of the magnetic wave vector which occurs
at the transition to the AFI. 
In addition, the insulating paramagnet is characterized
by magnetic correlations which are shorter ranged than those of
the paramagnetic metal. 
Like many single-band Hubbard calculations, the doped copper oxides
which eventually become high temperature superconductors display none of 
these anomalous features. 
  
The most obvious potential cause for the anomalous 
short range correlations in the PI phase is
that conventional exchange interactions between spins in 
insulating $\rm (V_{0.97}Cr_{0.03})_2O_3$ lead to a frustrated Heisenberg 
model which is unable to develop long range order at T$\sim$J/k$_B$. 
Solid state chemistry, however, speaks against this possibility because only 
A and B types of nearest neighbors (see Fig.~1)
with direct cation-cation overlap have appreciable exchange 
interactions in $\rm V_2O_3$\cite{castl} and a Heisenberg model with 
only these interactions is not frustrated\cite{Bertaut}. 

A more likely explanation of our data is based on work of Castellani et al. who
established, almost twenty years ago, how covalent bonding between doubly degenerate $3d$ orbitals control the sign and magnitude of
spin-spin interactions in $\rm V_2O_3$ \cite{castl}. The basic idea is that
the magnetically active electron on a single vanadium ion can exist in one
of two degenerate orbitals. Even so, the exchange integrals, which determine
the coupling between spins on neighboring ions, depend strongly on which 
orbitals are occupied. Indeed, when orbitals on neighboring ions are
 orthogonal, the resulting spin coupling is ferromagnetic, while
if they are not, the coupling tends to be antiferromagnetic. The net
Hamiltonian for insulating V$_2$O$_3$ then involves orbital as well
as spin degrees of freedom at each site, where the spin-orbit coupling is 
not a bilinear coupling on a single site, as 
in conventional magnets, but actually involves the relative 
spin and orbital occupancies on neighboring sites\cite{castl}. Castellani and
coworkers examined many possible orbital configurations for V$_2$O$_3$ and
concluded that the most likely is that represented by the colors used
to locate the V ions in Fig.~1. Here, the
magnetic unit cell is doubled in the basal planes because the orbital
ordering doubles the unit cell also. On warming through T$_N$, we are 
left with disorder in both spin and orbital occupancies. The magnetic  short
range order in the paramagnetic phase then depends on whether we are looking 
at frequencies (a) large or (b) small compared to the relaxation rate,
$\Gamma_{orb}$, characterizing
the orbital fluctuations. When (a) $\omega \gg \Gamma_{orb}$,
the interactions with
the neighbors will be similar to those below T$_N$, and we expect
to see heavily damped renditions of the spin waves seen below T$_N$.
On the other hand, if (b) $\omega \ll \Gamma_{orb}$,
one can average over the 
orbital degrees of freedom and obtain an effective spin Hamiltonian 
where all couplings are antiferromagnetic. 

We now consider whether limit (a) or (b) is more appropriate for our
$\rm (V_{0.97}Cr_{0.03})_2O_3$ sample. Obviously, 
because we see no remnants of the low-temperature spin
waves, description (a) cannot apply. We can check whether (b) applies,
especially in the detailed sense of its prediction that all interactions
are antiferromagnetic, by rewriting the observed structure factor
${\cal S}({\bf Q},\omega)$ as the Fourier transform of the two-spin correlation 
function in real space and then checking the signs of the near-neighbor
correlations. A remarkably good fit
($\chi^2=1.5$) to the experimental ${\cal S}({\bf Q},\omega)$\cite{3mevdata}
is obtained when we truncate the series to include correlations among
only four types of spin pairs (see inset of Fig.~1). More
specifically, we find that $\langle {\bf S}_0\cdot {\bf
S}_{[0,0,1/6+\delta]}\rangle^A =0.6(3)$, $\langle {\bf S}_0\cdot {\bf
S}_{[1/3,2/3,\overline{\delta}]}\rangle^B =-0.19(8)$, $\langle {\bf
S}_0\cdot {\bf S}_{[2/3,1/3,\delta-1/6]}\rangle^C =0.18(8)$ and
$\langle {\bf S}_0\cdot {\bf S}_{[2/3,1/3,1/6]}\rangle^D =-0.09(3)$,
where $\delta=0.026$ and we normalized so that
$\langle {\bf S}_0\cdot {\bf S}_0\rangle =1$. The curve through the
205K data in Fig.~3(d) was calculated using this model. Thus, 
measurable correlations are not all antiferromagnetic
and we  do not have a state (b) with complete
orbital disorder, i.e.\  with 
$\Gamma_{orb}/\omega \rightarrow \infty$.
Simultaneously, though, the antiferromagnetic correlations are very short-ranged
in spite of the fact that the temperature and measuring frequencies are below
the exchange constants characterizing the AFI phase\cite{wordr}.
The most probable cause is that
while $k_B$T and $\hbar \omega$ are somewhat lower than
$\Gamma_{orb}$, they are close enough to $\Gamma_{orb}$ that the 
spin couplings, fluctuating with the orbital occupancies, have effectively
random signs which give rise to magnetic frustration.
In other words, we are proposing that the orbital fluctuation
rate is of order $k_B$T$\approx$20meV. 

Finally, what happens in the metallic state? Here we can no longer 
speak about local orbital and spin degrees of freedom, but it is still
possible to make Wannier projections of analogous objects from the 
band states. Our finding that the magnetic correlations in the metallic state 
are much more similar to those in the paramagnetic insulator than
in the antiferromagnetic insulator corroborates the assignment by  
Takigawa et al\cite{takig} of the nuclear magnetic 
relaxation primarily to orbital rather than spin 
fluctuations in metallic V$_2$O$_3$. Even so, the orbital fluctuations may be
somewhat faster than in the PI because the antiferromagnetic correlations are 
further ranged (i.e. there are resolvably sharper peaks in ${\cal S}({\bf Q},\omega)$ 
for the PM phase) and the PM-AFI transition  appears more strongly
first-order than the PI-AFI transition\cite{wordr}. 

To summarize, we have discovered that the magnetic fluctuations in V$_2$O$_3$ 
switch dramatically as a function of temperature as the antiferromagnetic
insulating state is entered from either insulating or metallic paramagnetic
phases. They change in a much more modest fashion as a
function of doping and temperature in  the paramagnetic phase, be it metallic
or insulating. We conclude that the primary order parameter for the
AFI phase is orbital, and that orbital order drives the 
spin ordering not via a conventional spin-orbit interaction for
each V ion, but instead  via a long-range ordered  modulation of the exchange
constants coupling spins on different V ions.
The orbital ordering breaks translational symmetry, and is the orbital
analog of antiferromagnetism. This is to be contrasted with the celebrated
orbital order in the cubic uranium pnictides\cite{actin}, which while 
it leads to spectacular anisotropies, does not break 
translation symmetry and indeed can be traced to single-ion spin-orbit coupling.
Finally, our data represent strong evidence for the long-standing 
notion that orbital, charge, and spin degrees of freedom need to be
considered on an equal footing near the metal-insulator transition in 
generic transition metal oxides.

We gratefully acknowledge
discussions with S. K. Sinha, T. M. Rice, G. Kotliar, Q. M. Si, G. Sawatzky,
L. F. Mattheiss and R. J. Birgeneau. 
W.\ B.\ thanks the Aspen Center for Physics where part of the work
was performed.
Work at JHU was supported by the NSF through DM-9453362,
at ORNL by DOE under Contract No.\ DE-AC05-84OR21400,
at BNL by DOE under Contract No.\ DE-AC02-76CH00016 and
J. M. H. was supported on MISCON Grant DE-FG02-90ER45427.

\begin{figure}
\caption{(color) Phase diagram of $\rm V_{2-y}O_3$-$\rm (V_{1-x}Cr_x)_2O_3$ 
in the temperature-composition plane\protect\cite{phase_d,phase_d2}.
PI denotes the paramagnetic insulating phase, PM, 
the paramagnetic metallic phase,
AFI, the antiferromagnetic insulating phase, and SDW, the metallic 
spin-density-wave phase. The PM-PI phase boundary 
terminates at a critical point.
The color bars mark the samples and the temperature ranges explored 
in this work. 
Insert: spin and orbital orders in the AFI 
phase\protect\cite{moon,castl}. The two degenerate (in the single ion limit)
vanadium orbitals are
represented by red and blue respectively.
A-D label four kinds of near neighbor pairs.}
\end{figure}

\begin{figure}
\caption{(color) Contour maps of ${\cal S}({\bf Q},\omega )$ for ${\bf Q}$ along
the (10$\ell$) direction in (a) $\rm V_{1.973}O_3$ at T=1.4K,
(b) $\rm V_2O_3$ at T=200K, and (c) $\rm (V_{0.97}Cr_{0.03})_2O_3$ at T=205K. 
Intensity is indicated by the color bars in units of $\mu_B^2$/meV per
unit cell, which contains 6 formulas units. 
The {\bf Q} range covers a Brillouin zone, with nuclear
Bragg points (10$\overline{2}$) and (104) at the ends.
Resolutions in both {\bf Q} and $\omega$ are narrower than the
widths of features by at least a factor 2\protect\cite{longp,newb} 
(see Fig. 3(b) and (d) for a few examples).}
\end{figure}

\begin{figure}
\caption{(color) Constant $\hbar\omega=9$ meV scans along $(10\ell)$
and constant $\hbar\omega=12$ meV scans along (0.5,0.5,$\ell$)
for $\rm V_{2-y}O_3$ ((a)-(b)) and
$\rm (V_{0.97}Cr_{0.03})_2O_3$ ((c)-(d))
inside (open symbols) and outside (filled symbols) the AFI phase.
$(10\ell)$ scans look similar over explored $\hbar\omega$-range (See Fig.~2).
The horizontal bars indicate the FWHM of 
the projection of the resolution function on the scan direction.
Peaks in (a) and (c) are resolution limited.
We use $E_f=13.7$ meV. Collimations are 60'-40'-40'-80', 60'-40'-80'-80' 
and 60'-40'-40'-60' at NIST for (a), (b) and (d) respectively;
(c) shows data from HFIR with 50'-40'-40'-70'.}
\end{figure}

\begin{figure} 
\caption{(color) Temperature variations of the neutron scattering 
intensity from $(10\ell )$-type and
$(0.5,0.5,\ell )$-type magnetic fluctuations in pure 
V$_2$O$_3$ ((a)-(b)) and (V$_{0.97}$Cr$_{0.03}$)$_2$O$_3$ ((c)-(d)). 
$(10\ell )$-type fluctuations at $\hbar\omega=9$ meV were probed with 
{\bf Q}=(1,0,2) in (a) and {\bf Q}=(1,0,0.8) in 
(c). $(0.5,0.5,\ell )$-type fluctuations at $\hbar\omega=12$ meV were 
monitored at {\bf Q}=(0.5,0.5,0.2) in (b), while we 
report the (0.5,0.5,$\ell$) $\bf Q$-integrated 
intensity in (d). 
The dashed lines indicate the location of the phase transition 
out of the AFI as 
determined by the disappearance of a magnetic Bragg peak at {\bf 
Q}=(0.5,0.5,0). All data were acquired upon increasing temperature from 
T=1.4K.} 
\end{figure} 

\end{document}